\documentclass[fleqn,10pt,twocolumn]{wlscirep}

\usepackage{dblfloatfix}

\title{On hybrid circuits exploiting thermistive properties of slime mould}

\author[1,2*]{Xavier Alexis Walter}
\author[2]{Ian Horsfield}
\author[1]{Richard Mayne}
\author[2]{Ioannis A. Ieropoulos}
\author[1]{Andrew Adamatzky}
\affil[1]{Unconventional Computing Center, University of the West of England, Bristol, BS16 1QY, United Kingdom}
\affil[2]{Bristol Bio-Energy Center, Universities of Bristol and of the West of England, Bristol, BS16 1QY, United Kingdom}
\affil[*]{xavier.walter@uwe.ac.uk}

\begin{abstract}
Slime mould  \textit{Physarum polycephalum}  is a single cell visible by unaided eye. Let the slime mould span two electrodes with a single protoplasmic tube: if the tube is heated to approximately $\approx$ $40^{\circ}$C, the electrical resistance of the protoplasmic tube increases from $\approx$ 3 M$\Omega$ to $\approx$ 10'000 M$\Omega$. The organism's resistance is not proportional nor correlated to the temperature of its environment. Slime mould can therefore not be considered as a thermistor but rather as a thermic switch. We employ the \textit{P. polycephalum} thermic switch to prototype hybrid electrical analog summator, NAND gates, and cascade the gates into Flip-Flop latch. Computing operations performed on this bio-hybrid computing circuitry feature high repeatability, reproducibility and comparably low propagation delays.
\end{abstract}
\begin{document}

\flushbottom
\maketitle
%
%
\thispagestyle{empty}

\section*{Introduction}

Slime mould \emph{Physarum polycephalum} is a  large single celled organism ~\cite{stephenson1994myxomycetes} whose amorphous body is able to form complex, optimised networks of protoplasmic tubules between spatially distributed nutrient sources. It has been demonstrated that the organism's natural foraging behaviour may be characterised as distributed sensing, concurrent information processing, parallel computation and decentralized actuation~\cite{adamatzky2010physarum,adamatzkyAdvancesPhysarum}. The ease of culturing and experimenting with \emph{P. polycephalum} makes this slime mould  an ideal substrate for real-world implementations of unconventional sensing and computing devices~\cite{adamatzky2010physarum}.  A range of hybrid electronic devices have recently  been implemented as experimental working prototypes. They include self-routing and self-repairing wires \cite{adamatzky2013physarumwire}, electronic oscillators \cite{adamatzky2014slimeoscillator}, chemical sensor \cite{whiting2014towards}, tactical sensor \cite{adamatzky2013slime},  low pass filter \cite{whiting2015transfer},  colour sensor \cite{adamatzky2013towards}, memristor \cite{gale2013slime, tarabella2015hybrid}, robot controllers \cite{tsuda2006robot,galeAndroid}, opto-electronic logical gates \cite{mayne2015slimegates}, electrical oscillation frequency logical gates \cite{whiting2014slimefrequency}, FPGA co-processor \cite{mayne2015towardsFPGA}, Shottky diode \cite{cifarelli2014non} and transistor \cite{tarabella2015hybrid}.

Several challenges still remain towards fabricating functionally useful slime mould computing devices.  For example, the live substrate's inherent variability renders cascading of various varieties of logical gate too unreliable for constructing complex computing circuitry. We detail here the creation of laboratory prototypes of functional, reliable electrically-coupled discrete and sequential logical devices that utilise slime mould as circuit elements. This was achieved through characterising bioelectrical output of slime mould protoplasmic tubes in response to an insulting stimulus --- localised heating. This rationale was based on the principle that the organism is able to isolate portions of its body (transiently and permanently) in response to harmful stimuli by ceasing the rhythmic movements of cytoplasm through certain anatomical regions --- which usually serves to distribute the contents of the cell and contribute to the production of motive force --- hence isolating them from metabolically active regions of the cell. Isolated protoplasmic tubes have a significantly higher electrical resistance than their unaltered counterparts \cite{mayne2015toward}, which provides a basis for distinguishing between logical states when monitoring bioelectrical activity.

\begin{figure}[t]
\centering
\includegraphics[width=\linewidth]{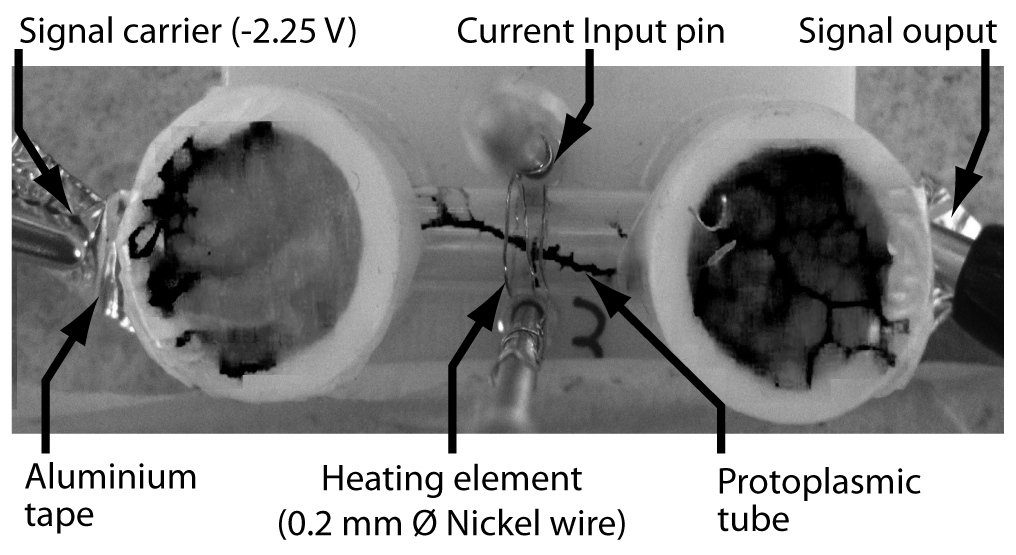}
\caption{Illustration of a H-tube (\textit{Physarum} heat-switch). A glass tube connects two 3D printed vessels. Each vessel hosts the slime mould. The slime mould connects spans two vessels with its protoplasmic tube.}
\label{fig:h-tube}
\end{figure}

\begin{figure*}[h]
\centering
\includegraphics[width=\textwidth, height=\textheight, keepaspectratio]{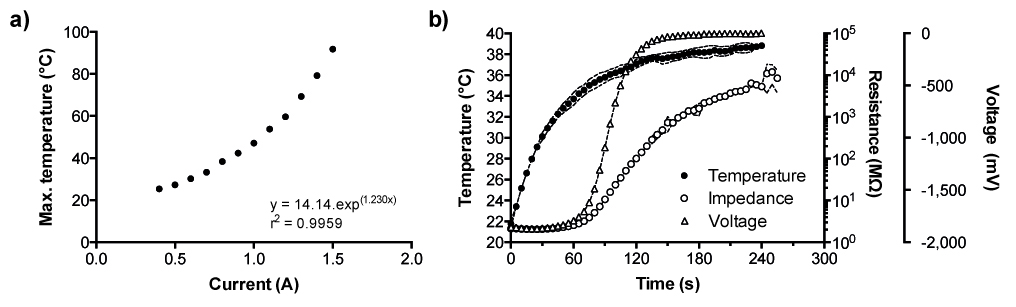}
\caption{Heat response behaviour of H-tubes. \textbf{a)} Dependence of the maximum temperature of a glass tube on the applied current. \textbf{b)} Example of the increase of resistance of a H-tube in response to heat generated by a 0.8 A current through the heating element. Dashed lines represent standard deviation ($n=3$).}
\label{fig:H-tube_parameter}
\end{figure*}

\begin{figure*}[!b]
\centering
\includegraphics[width=\textwidth, height=\textheight, keepaspectratio]{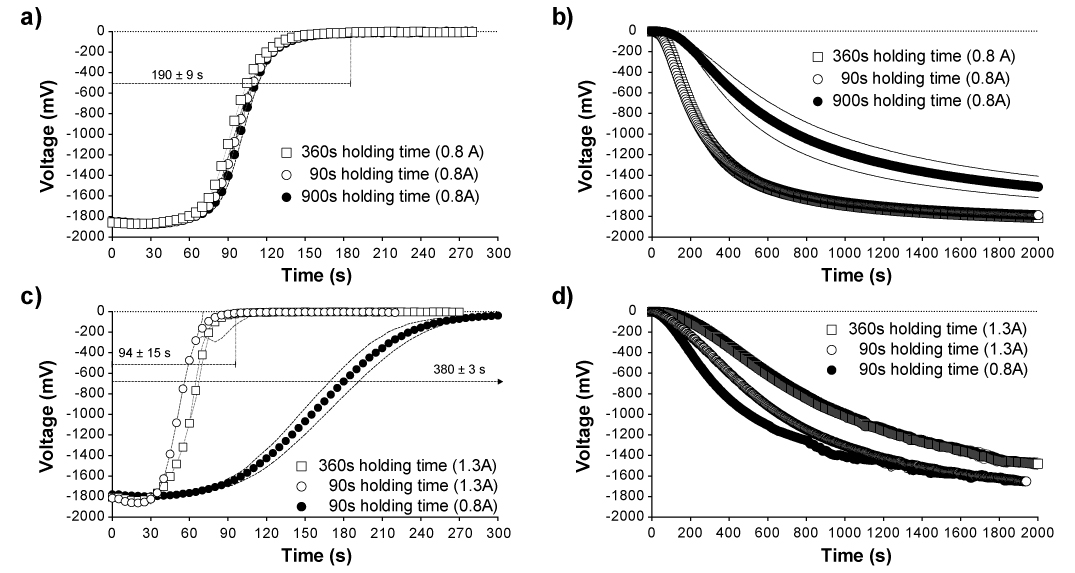}
\caption{Response time and reforming time of H-tube depending on the applied current and the holding time. \textbf{a)} Response time of H-tube 1 under 0.8 A current. \textbf{b)} H-tube 1 reforming times (0.8 A).  \textbf{c)} Response time of H-tube 2 under 1.3 A current. \textbf{d)} H-tube 2 reforming times (1.3 A). Shown data are averages and all dashed line stands for the standard deviation ($n=3$).}
\label{fig:Holding_short}
\end{figure*}

\begin{figure*}[h]
\centering
\includegraphics[width=\linewidth]{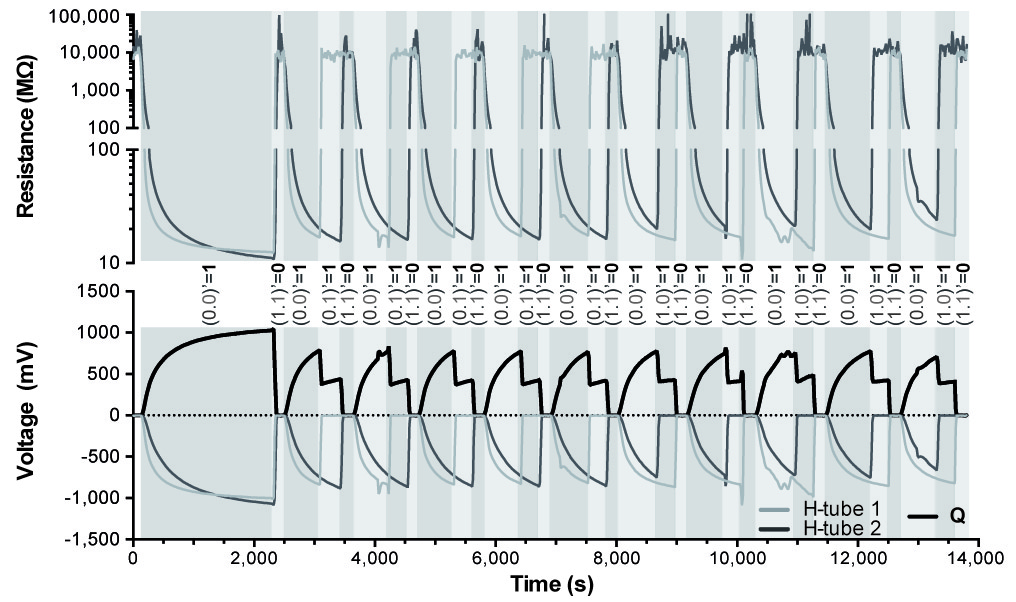}
\caption{NAND behaviour of the voltage signal output of 2 H-tubes (below 0 mV curves) connected to an OpAmp (above 0 mV curve). The coloured sections indicate a manual switching. Response to stimulation is interpreted as NAND logic and expressed as 
(H-tube1.H-tube2)=Output Q.}
\label{fig:Boolean}
\end{figure*}

\begin{table*}[ht]
\centering
  \begin{tabular}{ccc|c}
    \toprule
    Input current A & Input current B & Input current C & Output voltage Q\\
    \midrule
    \textbf{0} & \textbf{0} & \textbf{0} & \textbf{1 =$>$ \textbf{Q $\neq$ -31 mV}}\\
     -540.7 $\pm$ 58.5 mV & -850.9 $\pm$ 160.7 mV & -1074.5 $\pm$ 70.3 mV	& \ 159.9 $\pm$ 38.5 mV\\
    \hline
    \textbf{0} & \textbf{0} & \textbf{1} & \textbf{1 =$>$ \textbf{Q $\neq$ -31 mV}}\\
     620.4 $\pm$ 78.8 mV & -981.5. $\pm$ 296.0 mV & -2.5 $\pm$ 1.0 mV & -431.8 $\pm$ 75.1 mV\\
    \hline
   \textbf{0} & \textbf{1} & \textbf{0} & \textbf{1 =$>$ \textbf{Q $\neq$ -31 mV}}\\
     -599.7 $\pm$ 91.9 mV & -12.7 $\pm$ 1.9 mV & -1178.9 $\pm$ 73.1 mV & 404.9 $\pm$ 52.7 mV\\
    \hline
    \textbf{1} & \textbf{0} & \textbf{0} & \textbf{1 =$>$ \textbf{Q $\neq$ -31 mV}}\\
     -1.7 $\pm$ 0.4 mV & -939 $\pm$ 86.6 mV & -1148.3 $\pm$ 66. 3 mV & 307.6 $\pm$ 18.3 mV\\
    \hline
    \textbf{1} & \textbf{1} & \textbf{0} & \textbf{1 =$>$ \textbf{Q $\neq$ -31 mV}}\\
     -1.4 $\pm$ 0.2 mV & -11.1 $\pm$ 0.9 mV mV & -1204.4 $\pm$ 60.2 mV	& 568.1 $\pm$ 30.3 mV\\
    \hline
    \textbf{0} & \textbf{1} & \textbf{1} & \textbf{1 =$>$ \textbf{Q $\neq$ -31 mV}}\\
     -682.7 $\pm$ 76.0 mV & -9.9 $\pm$ 1.6 mV & -1.7 $\pm$ 0.1 mV	 & -204.7 $\pm$ 18.9 mV\\
    \hline
    \textbf{1} & \textbf{0} & \textbf{1} & \textbf{1 =$>$ \textbf{Q $\neq$ -31 mV}}\\
     -2.9 $\pm$ 1.7 mV & -932 $\pm$ 50.0 mV & -1.9 $\pm$ 0.3 mV & -265.0 $\pm$12.9 mV\\
    \hline
    \textbf{1} & \textbf{1} & \textbf{1} & \textbf{0 =$>$ \textbf{Q $\approx$ -31 mV}}\\
     -1.6 $\pm$ 0.3 mV & -9.7 $\pm$ 2.6 mV & -1.9 $\pm$ 0.3 mV & -33.9 $\pm$ 1.0 mV\\
    \bottomrule
  \end{tabular}
   \caption{Truth table of the 3 way analogue gate with the result interpreted as NAND logic. Due to a resistance shift of the 50 k$\Omega$ resistor in the second OpAmp adder, the 0 mV value was -31 $\pm$ 3 mV. Each condition of the truth table were done in triplicate, and all with the same H-tubes in a sequential manner limiting the holding time for a single H-tube to 15 minutes.}
   \label{tab:3-way}
\end{table*}

\begin{figure*}[h!t]
\centering
\includegraphics[width=\linewidth]{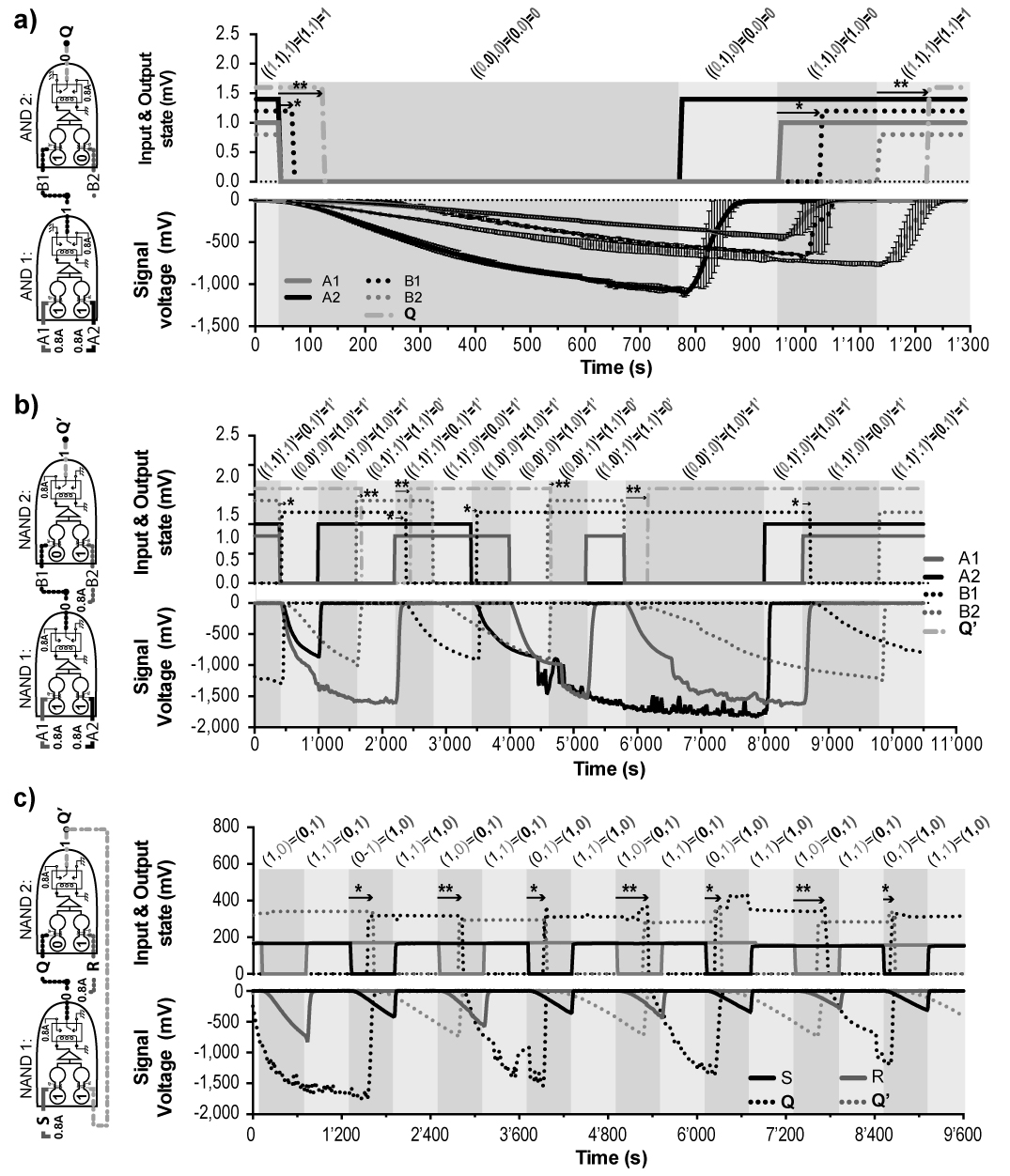}
\caption{Cascading logic gates illustrating the signal responses of the H-tubes to data inputs. \textbf{a)} Cascading AND gates. Error bars stand for the standard deviation ($n=3$).\textbf{b)} Cascading NAND gates. \textbf{c)} Low SR NAND Gate Latch. The coloured sections indicate manual switching. The $\ast$ represent the propagation delay of the first gate, and $\ast$$\ast$  the propagation delay of the second gate. In (\textbf{a,b}) the voltages of the gates (above 0 mV) are arbitrary values indicating when the coils were switched on. In (\textbf{c}) the curves above 0 mV are the measured voltages of the coils.} 
\label{fig:Gate}
\end{figure*}

\section*{Results and discussion}

\subsection*{Response of \emph{P.polycephalum} to heat}

The resistance of a protoplasmic tube  was found to increase when exposed to temperatures exceeding approximately 25$^{\circ}$C. The  glass tube temperatures of a H-tube device (Fig. \ref{fig:h-tube}) exceeding 70$^{\circ}$C (69.3 $\pm$ $0.6^{\circ}$C when heating element current was set to 1.3A) (Fig. Fig. \ref{fig:H-tube_parameter}a) were found cause irreparable damage to protoplasmic tubes. Heat input of approximately $40^{\circ}$C (0.8A) was found to be the most appropriate compromise between response speed whilst causing negligible damage to the organism (Fig. \ref{fig:H-tube_parameter}a), because it allowed the slime mould to reform promptly and be re-stimulated repeatedly over duty cycles extending for approximately 3000s. Figure \ref{fig:H-tube_parameter}b shows the voltage drop and resistance responses of a H-tube to heating  to this degree: response times were found to be 190 $\pm$ 9s (which is relatively rapid for \emph{Physarum} computing devices \cite{tsuda2004Gate,mayne2015slimegates,whiting2014slimefrequency}) and were independent to holding times at this value, although reforming times were found to be proportional to holding time (Fig. \ref{fig:Holding_short}). 

The resistance of a resting H-tube was found to be stable with an average of 2.27 $\pm$ 0.30 M$\Omega$ for the 20 mm distance covered by a protoplasmic tube. Under heat-stimulation, the resistance values of a H-tube fluctuated significantly with some peaking at 300,000 M$\Omega$. However, without taking into account these bursts, the average resistance was 9,450 $\pm$ 5,000 M$\Omega$, usually fluctuating between 7,000 and 15,000 M$\Omega$. Despite this variability, the values attained were sufficient to reliably distinguish between resting and stimulated states.

It was found, however, that for each H-tube to respond promptly to heat simulation, it was required to have had previous exposure to such a stimulus: this apparently adaptive response was conditioned into each experimental organism by exposing it to heat generated by a 0.9 A current until a positive response was measured, after which the temperature was maintained for 300 s in order to enhance reproducibility between different organisms. We considered this phase as an entrainment of \textit{P. polycephalum} towards heat stimulation. \textit{P. polycephalum} changes its resistance in response to heat. Therefore it can be regarded as a living thermistor. Its response to heat-stimulation only occurred if the temperature increased over a threshold specific to the physiological state of \textit{P. polycephalum} (thickness and number of protoplasmic tubes), however. Since \textit{P. polycephalum} resistance was not proportional to temperature, it cannot be considered as a `pure' thermistor, but as a thermic switch. The mechanisms behind the temperature sensitivity and the resistive variations are yet to be determined, although historical literature indicates the synthesis of heat-shock proteins in response to temperatures exceeding 32~$^o$C \cite{Wright1978} presumably allow the organism to acclimatise to such stimuli.

\subsection*{Boolean logical gates}

\subsubsection*{Two-ways analogue NAND logic}

To balance the response and reforming time in response to heat stimulation, the duty-cycle of the setup was as follows: resting condition for 10 min (no current), stimulated conditions for 8 min (5 + 3 mins; 0.8 A; Fig. \ref{fig:Setup}a). Under the applied running conditions, both of the tested setups (H-tube1 and H-tube2) demonstrated reproducible and stable response behaviour to heat stimulation (Fig. \ref{fig:Boolean}). With a reforming time of 10 min, the resistance reached by the H-tubes was of $\approx$ 25 M$\Omega$. The resistance of the H-tube under stimulation was of $\approx$ 10,000 M$\Omega$. 

Even though these resistances were high, the difference was sufficient for the behaviour to be interpreted as a NAND logic: outputs (voltage) are high when input (current) configuration is $\langle$0,0$\rangle$, $\langle$0,1$\rangle$, $\langle$1,0$\rangle$; the outputs are low when input (current) configuration is $\langle$1,1$\rangle$ (0 = no current, 1 = current applied to heating element). As it is an analogue setup, the high outputs comprise three levels: $\langle$0,0$\rangle$ condition results in the highest voltage output value; $\langle$0,1$\rangle$, $\langle$1,0$\rangle$ conditions have a voltage outputs comprised between high and low outputs, and are defined as medium outputs. However, because both H-tubes had comparable physiological states, their respective resistance responses were varying at similar levels. Therefore, the $\langle$0,1$\rangle$, and $\langle$1,0$\rangle$ conditions had equivalent voltage output Q and could not be deduced from the OpAmp output: 419 $\pm$ 7.5 mV 432 $\pm$ 23.5 mV, respectively. Nonetheless, results demonstrate that the H-tube response to heat could be interpreted as analogue NAND logic, with a duty cycle of 18 min whereby the response and reforming times were balanced (Fig. \ref{fig:Boolean}).

\subsubsection*{Three-ways analogue NAND logic}

Due to the limitations of the measuring equipment (c.a. +$/$- 2250mV) and to the need for a sufficient spread of various output voltages, the third H-tube was not connected to the same OpAmp as the two first ones: the two first H-tubes were connected to one OpAmp, the output of which was then connected to a second OpAmp in parallel to the third H-tube. However, this setup was equivalent to a three inputs OpAmp system adapted to the data logging apparatus (Fig. \ref{fig:Setup}b).

In order to have a specific output voltage for each logic combination, the H-tubes chosen for the three-way NAND gate had different physiological states (i.e. thickness and number of protoplasmic tubes; see resulting voltage from Input currents = 0 in Tab. \ref{tab:3-way}). Results confirmed that when the organism had different physiological states the voltage output levels were specific to each logical combination (Kruskal-Wallis test, p$<$ 0.0001; Tab. \ref{tab:3-way}). Indeed, each of the $\langle$0,0,1$\rangle$, $\langle$0,1,0$\rangle$, $\langle$1,0,0$\rangle$, $\langle$1,1,0$\rangle$, $\langle$0,1,1$\rangle$, $\langle$1,0,1$\rangle$ conditions had different medium voltage outputs Q (Tab. \ref{tab:3-way}). Therefore, the voltage outputs could be interpreted as analogue NAND logic with each inputs condition identifiable from the output voltage. Since different, each output could be routed to other components of a larger electronic setup.

\subsection*{Cascading logic gates}

\subsubsection*{AND to AND}

The relay of the first gate was connected open to the coil of the second gate (Fig. \ref{fig:Setup}c). In this context, when the voltage output of the first OpAmp was $\leq$ 10 mV the relay was switched to the closed position, thus heating the coil of one of the H-tubes in the cascading gate. Therefore, when the first gate was $\langle 1,1 \rangle$, the following H-tube of the second gate became 1 (Fig. \ref{fig:Gate}a).
When the three H-tubes were switched to Current Input = 0, the propagation delay, for the H-tube B1 (first $\ast$, Fig. \ref{fig:Gate}a) to switch to Current Input = 0, was of 18 $\pm$ 8 s. The propagation delay, for the output Q of the second AND gate (first $\ast$$\ast$, Fig. \ref{fig:Gate}a) to switch to Current Input = 0 after B1 switched to Current Input = 0, was of 62 $\pm$ 5 s. Thus, the overall propagation delay of the cascading AND gates was of 80 $\pm$ 5 s. The propagation delay, for the H-tube B1 (second $\ast$, Fig. \ref{fig:Gate}a) to switch to Current Input = 1, was of 62 s $\pm$ 25 s. The propagation delay, for the H-tube B1 (second $\ast$$\ast$, Fig. \ref{fig:Gate}a) to switch to Current Input = 1, was of 78 s $\pm$ 19 s. The data shown in Figure \ref{fig:Gate}a are averages of three succeeding duty cycles. Therefore, the initial part of the graph illustrates the robustness of the setup resilience. In the case of an AND cascading into AND gate, the propagation of the information essentially depends on the response time of \textit{P. polycephalum} to heat, whereas the reset of the setup is depending on the reforming time/speed of \textit{P. polycephalum}.

\subsubsection*{NAND to NAND}

In the case of a NAND cascading into a NAND, the relay of the first gate was connected in closed position to the coil of one of the H-tube in the cascading gate (Fig. \ref{fig:Setup}c). This implies that the coil was always under 0.8 A, and that when the voltage output of the first OpAmp was $\leq$ 10 mV the relay was switch into open position. Set as such, the gate acted as a NAND: the coil of the cascading H-tube (B1) was a 0 only when the first gate had both inputs as 1.

A NAND gate cascading into a NAND gate implies that most of the time of the cascading input of the second NAND gate is under heat stimulation (B1, Fig. \ref{fig:Gate}b). The propagation delay of Q$'$ was therefore longer when the computation was only depending on the reforming time of B1 (c.a. B2 was 1; second $\ast$ in Fig. \ref{fig:Gate}b). After 85 min of heat stimulation the reforming speed of B1, even though 30\% slower than at the start of the experiment, was yet sufficiently quick to be employed for computational purposes. 

In this experiment the Undefined State of the Low SR NAND Gate Latch was not investigated, but all other states were with the following sequence, expressed as (S,R)=(Q,Q$'$): (1,0)=(0,1), (1,1)=(0,1), (0,1)=(1,0) and (1,1)=(1,0). All these conditions were maintained during 10 minutes. The propagation delay to pass from Reset (Q,Q$'$ = 0,1) to Set state (Q,Q$'$ = 1,0) was 240 $\pm$ 40 s ($\ast$ in Fig. \ref{fig:Gate}c). This propagation delay mainly depended on the reforming speed of S, and the response time of Q. The propagation delay to pass from Set state to Reset state was 385 $\pm$ 55 s ($\ast$$\ast$ in Fig.\ref{fig:Gate}c). This propagation delay mainly depended on the reforming speed of R, and the response time of Q$'$. 

As discussed previously (Fig. \ref{fig:Holding_short}), the main limiting factor to a quicker propagation delay, in these setups, is the reforming speed of \textit{P. polycephalum} after a heat stimulation. This reforming speed could be faster with a lower current input (i.e. lower temperatures), but then it would be the response time to heat that would become limiting. Another approach that could be pursued would be to have thinner (i.e. smaller diameter glass tube) and shorter (i.e. length of the glass tube) protoplasmic tubes.

\section*{Conclusions}

The presented results demonstrate direct cascading setups employing the biological response of \textit{P. polycephalum} to stimulation. In the present case, it was the response to heat that was exploited to build the hybrid gates with living slime mould. The full duty cycle, from the heat response to reforming, was of $\pm$ 40-60 min. Besides, results have shown that in order to implement this heat stimulation as a data input in slime mould hybrid gates, the reforming time can be shorten by $\approx$ 75\% for it to be balanced with the response time to stimulation. To exploit this heat response for implementing Boolean logic, \textit{P. polycephalum} was employed as ``heat-switch'' connected in pair to a summing amplifier. Furthermore, as the input and output voltage levels were not compatible for cascading the gates, a comparator and a relay were employed. The resulting gate comprised two individuals of \textit{P. polycephalum}, a summing amplifier, a comparator and a relay.

The present study reports AND to AND and NAND to NAND cascading gates, either as linear cascades or as a low SR NAND gate Flip-Flop latch. In the latter configuration, the propagation delays were of 240-380 $\pm$ 55 s. These gates were shown to be reliable and resilient, thus leading to the run of multiple duty cycles with the same setups, whilst producing reproducible results from one duty cycle to another. Our observation of an adaptive response which sensitises the organism to the application of heat stimuli is consistent with contemporary knowledge of the organism’s ability to periodic stimuli \cite{Persin2009Memristive}. Therefore, the necessary entrainment stage of these devices’ operation could be interpreted as analogous to memory --- like loading a program into the organism’s prior to it being run --- but would require further investigation to decipher if this "adaptive memory" is not the simple consequence of any passive metabolite accumulation in the protoplasmic tubes.

\section*{Methods}

\subsection*{Culturing \textit{Physarum polycephalum}}

\textit{P. polycephalum} was kept in constant plasmodial growth phase by regular sub-culturing, 1-2 times per week, on non-nutritive agar plates (2\%). These agar-containing Petri dishes comprised five blobs ($\leq$ 5 mm diameter) of sterile porridge as food sources and were kept in dark at 22 $\pm$ 0.5$^{\circ}$C. When required, a subsample was employed to inoculate, under sterile conditions, sterile vessels specifically designed to apply heat to \textit{P. polycephalum} protoplasmic tubes and measure its response (H-Tubes; see next paragraph). The porridge was prepared by adding 20 mL of deionised autoclaved water to 20 g of three-time autoclaved oat flakes. Once the mixture had set thickly, it was autoclaved once more. This thick texture allowed cutting small squares of the needed size and facilitate re-spiking/inoculation.  

\subsection*{Experimental environment} 

An enclosed, heat-resistant, electrically insulative environment was designed to enable the growth of plasmodial tubules between two electrodes. This \textit{P. polycephalum} heat switch device (H-tube) (Fig. \ref{fig:h-tube}) comprised of two 10mm diameter 3D-printed wells, each holding 500 $\mu$l of sterile agar  whose surface was levelled to the bottom of a 6 mm external diameter glass tube which connected the two wells. The glass tubes were 20 $\pm$ 0.3 mm in length and had a coil of 45 x 0.2mm nickel wire wrapped about their centre, which functioned as a heating element.
Pieces of sterile porridge were added in both wells and one of the two was inoculated with \textit{P. polycephalum}. Both wells were then sealed with paraffin film. The bottom of each well was laid with aluminium tape, one end of which was accessible from the outside of the device, to serve as the signal carrier. Hence, when the organism propagated from one well to the other, it formed at least one protoplasmic tube between the wells, allowing for electrical measurements to be made non-invasively along the length of the organism. 
Measurements were performed with a PicoLog ADC-24 24-bit resolution data logger (Pico Technology, UK): input channels were set to $\pm$ 2250mV recording limits and were ground referenced. As there was significant discrepancy between the device's internal resistance (1M$\Omega$) and that of the protoplasmic tubes ($\approx$ 3 M$\Omega$), a voltage follower was adapted into the signal path.

\subsection*{Response of \textit{P. polycephalum} to heat}

\begin{figure*}[!h]
\centering
\includegraphics[width=0.9\textwidth, height=0.9\textheight, keepaspectratio]{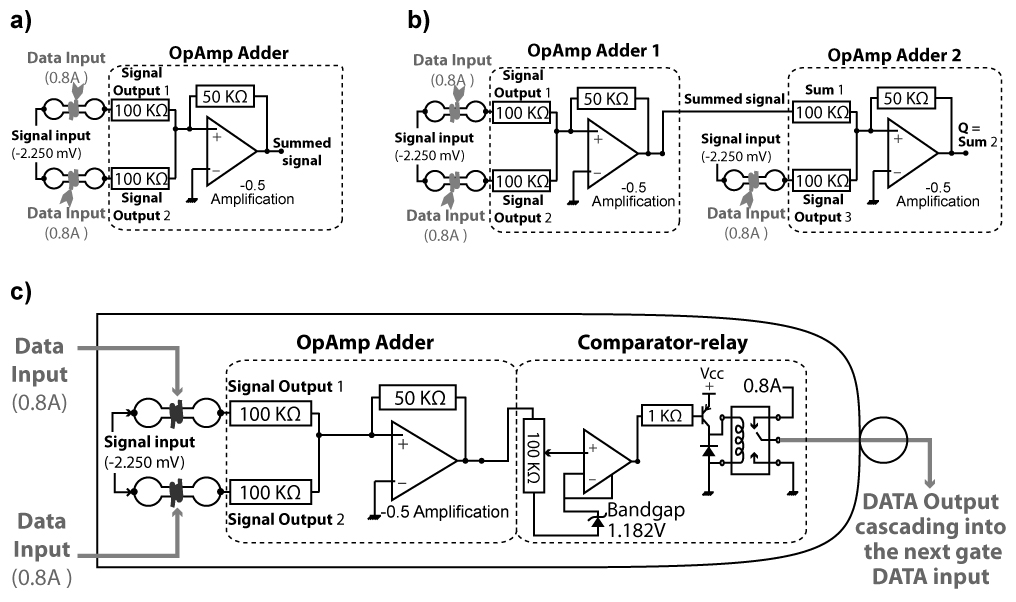}
\caption{Circuit diagrams of hybrid H-tube/electronic circuitry.  \textbf{a)} Summing adder comprising two H-tubes feeding into a summing amplifier. \textbf{b)} 3-way summing amplifier circuitry.  \textbf{c)} Bio-logic gate comprised of summing amplifier connected to a comparator-relay unit.}
\label{fig:Setup}
\end{figure*}

In preliminary experiments, H-tubes were exposed to a range of heat levels --- adjusted by varying the current passing through the nickel wire --- for varying periods of time whilst concurrently measuring bioelectrical activity of the constituent protoplasmic tube, in order to assess the organism's electrical responses to the stimulus. The protoplasmic tube was loaded with -2.25 V and the voltage drop across the organism was monitored in response to heat changes; temperature of the glass tube was concurrently measured using a Fluke 87V multimeter. This methodology was chosen in order to assess the optimum temperature for substantially affecting functionality of the protoplasmic tube --- hence causing its resistance to raise to such a degree that the measured voltage drops significantly --- that it reacts rapidly, but is not so damaged that it could not revert to a normal state once the insulting stimulus has been removed.

\subsection*{Hybrid circuitry} 

\subsubsection*{Design}

The input states of the gates were set via the use of heater coils placed around the glass tubes the \textit{P. polycephalum} were grown in (Fig. \ref{fig:h-tube}). A negative voltage of -2.25 V was applied to one end of these tubes which formed potential dividers in conjunction with 10 M$\Omega$ resistors. The resultant voltages derived were immediately buffered via unity gain amplifiers to remove any possible errors due to the impedance loading of the instrumentation (not shown on the circuit diagrams in Fig. \ref{fig:Setup}). The buffered signals were then fed into a summing amplifier (OpAmp) with the gain set to minus a half \textit{of the summed input voltages} (Fig. \ref{fig:Setup}a). Due to the high resistivity of protoplasmic tubes, it was not possible to employ \textit{P. polycephalum} as a data carrier. Hence, each device consisted of two electrical circuits: the first consisted of a voltage supply for loading the protoplasmic tubes with a constant -2.25 V and the second was the current source for the heating elements, which comprised the devices' inputs; this prevented the devices being directly cascaded. To overcome this problem, the signal from the summing amplifier was passed onto a simple comparator circuit which controlled the heater supply of a \textit{P. polycephalum} output tube via a transistor driven relay circuit, depending on the applied voltage and the threshold set via a potentiometer (Fig. \ref{fig:Setup}c). The following devices were evaluated: OpAmp adder (summing amplifier), three-input OpAmp adder and variable-configuration logical gate comprised of an OpAmp adder and comparator relay, the circuits for which are shown in Figure \ref{fig:Setup}.

\subsubsection*{Analogue logical gates}

A two-input NAND gate was constructed using the device detailed in Figure \ref{fig:Setup}a. Its functionality was as follows: outputs (current) are high and low when input (voltage) configuration is $\langle 0,0 \rangle$, $\langle 1,1 \rangle$, respectively, as with a conventional digital NAND gate. When only one input is high, however, the output voltage is a medium value between high and low. If logic levels are interpreted as high, i.e. above a certain threshold lying between the low and medium states, the output of the device is essentially equal to its digital equivalent, such that the output configuration is $\neg$(A.B).
Evaluation of this gate was performed by setting the inputs to $\langle 0,0 \rangle$, $\langle 0,1 \rangle$, $\langle 1,1 \rangle$ five times for 10, 5 and 3 minutes, respectively, without interruption in a single H-tube; the experiment was then repeated a further five times with the 5 minute phase in configuration $\langle 1,0 \rangle$.
Consequently, a three-input NAND gate was evaluated utilising the device shown in Figure \ref{fig:Setup}b. All possible combinations of input configuration were evaluated three times. 

\subsubsection*{Cascading logic gates}

The logical gate shown in Figure \ref{fig:Setup}c contained a comparator relay component for controlling heating elements, hence providing a coupling mechanism for cascading multiple gates. The logic level can be manually set as a voltage threshold with the integrated potentiometer. When viewed as a single logical gate, this device can be set up to function as AND or OR gates, plus their negated versions --- NAND, NOR --- via the switch for the heater supply: depending on whether the normally open or normally closed contacts of the relay were used, the NOT function was implemented. The following sequential logical circuits were constructed through cascading two gates:
\begin{itemize}
\item AND to AND, such that ((A1.A2).B2) = (B1.B2) = Q.
\item NAND to NAND, such that ((A1.A2)$'$.B2)$'$ = (B1.B2)$'$= Q$'$.
\item Low SR flip-flop (bistable) utilising two NAND gates, such that (S.R)=(Q.Q$'$).
\end{itemize}

Their evaluation was performed by subjecting them to repeated duty cycles.

\section*{Acknowledgements}

The authors gratefully acknowledge the EU Commission seventh Framework Program for funding the PhyChip Project (FP7-ICT-2011-8/316366).

\section*{Author contributions statement}

AA developed the original concept of employing \textit{P. polycephalum} as bio-logic gates. XAW developed the concept --- and designed the adequate prototype --- of employing the response of \textit{P. polycephalum} to heat as the data input. XAW, AA and IH have participated to the experimental designs. XAW conducted the experiments, collected data and produced the figures, and wrote the draft manuscript. All authors were involved in the data analysis and interpretation. All authors edited and approved the final manuscript.
 
\section*{Additional information}

\textbf{The authors declare that they have no competing interests.}


\begin{thebibliography}{}
\expandafter\ifx\csname url\endcsname\relax
  \def\url#1{\texttt{#1}}\fi
\expandafter\ifx\csname urlprefix\endcsname\relax\def\urlprefix{URL }\fi
\providecommand{\bibinfo}[2]{#2}
\providecommand{\eprint}[2][]{\url{#2}}

\end{thebibliography}


\begin{thebibliography}{100}

\bibitem{adamatzky2010physarum}
Andrew Adamatzky.
\newblock  Physarum machines: computers from slime mould, volume~74.
\newblock {\em World Scientific}, 2010.


\bibitem{adamatzky2013physarumwire}
Andrew Adamatzky.
\newblock Physarum wires: Self-growing self-repairing smart wires made from
  slime mould.
\newblock {\em Biomedical Engineering Letters}, 3(4):232--241, 2013.

\bibitem{adamatzky2013slime}
Andrew Adamatzky.
\newblock Slime mould tactile sensor.
\newblock {\em Sensors and actuators B: chemical}, 188:38--44, 2013.

\bibitem{adamatzky2013towards}
Andrew Adamatzky.
\newblock Towards slime mould colour sensor: {R}ecognition of colours by
  {P}hysarum polycephalum.
\newblock {\em Organic electronics}, 14(12):3355--3361, 2013.

\bibitem{adamatzky2014slimeoscillator}
Andrew Adamatzky.
\newblock Slime mould electronic oscillators.
\newblock {\em Microelectronic Engineering}, 124:58--65, 2014.

\bibitem{adamatzky2014tactile}
Andrew Adamatzky.
\newblock Tactile bristle sensors made with slime mold.
\newblock {\em Sensors Journal, IEEE}, 14(2):324--332, 2014.

\bibitem{adamatzky2015alife}
Andrew Adamatzky.
\newblock A would-be nervous system made from a slime mold.
\newblock {\em Artificial Life}, 21(1):73--91, 2015.

\bibitem{adamatzkyAdvancesPhysarum}
Andrew Adamatzky, editor.
\newblock {\em Advances in Physarum machines: Sensing and computing with slime
  mould}.
\newblock Springer, 2016.


\bibitem{cifarelli2014non}
Angelica Cifarelli, Alice Dimonte, Tatiana Berzina, and Victor Erokhin.
\newblock Non-linear bioelectronic element: Schottky effect and
  electrochemistry.
\newblock {\em International Journal of Unconventional Computing},
  10(5-6):375--379, 2014.


\bibitem{galeAndroid}
Ella Gale and Andrew Adamatzky.
\newblock Translating slime mould responses: A novel way to present data to the
  public.
\newblock In Andrew Adamatzky, editor, {\em Advances in Physarum Machines},
  Heidelberg, 2016. Springer.

\bibitem{gale2013slime}
Ella Gale, Andrew Adamatzky, and Ben de~Lacy~Costello.
\newblock Slime mould memristors.
\newblock {\em BioNanoScience}, 5(1):1--8, 2013.

\bibitem{mayne2015slimegates}
Richard Mayne and Andrew Adamatzky.
\newblock Slime mould foraging behaviour as optically coupled logical
  operations.
\newblock {\em International Journal of General Systems}, 44(3):305--313, 2015.


\bibitem{mayne2015toward}
Richard Mayne and Andrew Adamatzky.
\newblock Toward hybrid nanostructure-slime mould devices.
\newblock {\em Nano LIFE}, 5(01):1450007, 2015.

\bibitem{mayne2015towardsFPGA}
Richard Mayne, Michail-Antisthenis Tsompanas, Georgios~Ch Sirakoulis, and
  Andrew Adamatzky.
\newblock Towards a slime mould-FPGA interface.
\newblock {\em Biomedical Engineering Letters}, 5(1):51--57, 2015.

\bibitem{Persin2009Memristive}
Pershin, Yuriy V., Steven La Fontaine,  and Massimiliano Di Ventra.
\newblock Memristive model of amoeba's learning.
\newblock {\em Physical Review E}, 80:021926, 2010.

\bibitem{stephenson1994myxomycetes}
Steven~L Stephenson, Henry Stempen, and Ian Hall.
\newblock {\em Myxomycetes: a handbook of slime molds}.
\newblock Timber Press Portland, Oregon, 1994

\bibitem{tarabella2015hybrid}
Giuseppe Tarabella, Pasquale D'Angelo, Angelica Cifarelli, Alice Dimonte,
  Agostino Romeo, Tatiana Berzina, Victor Erokhin, and Salvatore Iannotta.
\newblock A hybrid living/organic electrochemical transistor based on the
  Physarum polycephalum cell endowed with both sensing and memristive
  properties.
\newblock {\em Chemical Science}, 6(5):2859--2868, 2015.

\bibitem{tsuda2004Gate}
Soichiro Tsuda, Masashi Aono, and Yukio-Pegio Gunji.
\newblock Robust and emergent Physarum logical-computing.
\newblock {\em BioSyatems}, 73:45--55, 2004.

\bibitem{tsuda2006robot}
Soichiro Tsuda, Klaus-Peter Zauner, and Yukio-Pegio Gunji.
\newblock Robot control: From silicon circuitry to cells.
\newblock {\em Biologically Inspired Approaches to Advanced Information
  Technology}, pages 20--32, 2006.

\bibitem{whiting2014slimefrequency}
James~GH Whiting, Ben~PJ de~Lacy~Costello, and Andrew Adamatzky.
\newblock Slime mould logic gates based on frequency changes of electrical
  potential oscillation.
\newblock {\em Biosystems}, 124:21--25, 2014.

\bibitem{whiting2014towards}
James~GH Whiting, Ben~PJ de~Lacy~Costello, and Andrew Adamatzky.
\newblock Towards slime mould chemical sensor: {M}apping chemical inputs onto
  electrical potential dynamics of {P}hysarum polycephalum.
\newblock {\em Sensors and Actuators B: Chemical}, 191:844--853, 2014.

\bibitem{whiting2015transfer}
James~GH Whiting, Ben~PJ de~Lacy~Costello, and Andrew Adamatzky.
\newblock Transfer function of protoplasmic tubes of Physarum polycephalum.
\newblock {\em Biosystems}, 128:48--51, 2015.

\bibitem{Wright1978}
Michelle Wright, and Yvette Tollon.
\newblock Heat sensitive factor necessary for mitosis onset in \emph{Physarum polycephalum} (temperature shift/heat shock/cycloheximide/ts mutant)/
\newblock {\em Molecular and General Genetics}, 163:91--99, 1978.


\end{thebibliography}
\end{document}